\documentstyle[prl,aps,twocolumn]{revtex}
\large
\begin{document}
\draft
\twocolumn[\hsize\textwidth\columnwidth\hsize\csname @twocolumnfalse\endcsname

\title{Critical Fluctuations and Disorder at the
Vortex Liquid to Crystal Transition\\
in Type-II Superconductors}
\author{M.A.Moore $^{(1)}$ and T.J.Newman $^{(2,3)}$}
\address{$^{(1)}$Department of Physics, University of Manchester,
Manchester, M13 9PL, U.K.\\
$^{(2)}$Theoretical Physics, University of Oxford,
Oxford, OX1 3NP, U.K.\\
$^{(3)}$Institut f\"ur Theoretische Physik, Universit\"at zu K\"oln,
D-50937 K\"oln, Germany \cite{pa}\\}
\date{April 6, 1995}
\maketitle
\begin{abstract}
We present a functional renormalization group (FRG) analysis of a
Landau-Ginzburg model of type-II superconductors (generalized to
$n/2$ complex fields) in a magnetic field, both for a pure system,
and in the presence of quenched random impurities. Our analysis is
based on a previous FRG treatment of the pure case $\lbrack $ E.Br\'ezin et.
al., Phys.  Rev. B,{\bf 31}, 7124 (1985) $\rbrack$ which is an expansion in
$\epsilon = 6-d$.  If the coupling functions are restricted to the
space of functions with non-zero support only at reciprocal lattice
vectors corresponding to the Abrikosov lattice, we find a stable FRG
fixed point in the presence of disorder for $1<n<4$, identical to
that of the disordered $O(n)$ model in $d-2$ dimensions. The pure
system has a stable fixed point only for $n>4$ and so the physical
case ($n = 2$) is likely to have a first order transition. We
speculate that the recent experimental findings that disorder
removes the apparent first order transition are consistent with
these calculations.
\end{abstract}
]

\narrowtext

In 1957 Abrikosov \cite{ab} demonstrated the existence of the `mixed
phase' in type-II superconductors wherein the external magnetic field
penetrates the sample in the form of a triangular flux line array
which now bears his name -- the Abrikosov lattice. This work was based
on a mean-field analysis of a Landau-Ginzburg formulation of the
problem, and the transition from the normal phase to the mixed phase
was found to be continuous.  The natural question which then arises
is: how do fluctuations affect the nature of this phase transition?
This question was finally addressed by Br\'ezin, Nelson and Thiaville
(BNT) in the form of a functional renormalization group (FRG) analysis
of the Landau-Ginzburg theory \cite{bnt}.  Their conclusion, based on
a numerical analysis of the (functional) flow equation, was that
fluctuations drive the transition from continuous to first order in
spatial dimension $d<6$.

Our original motivation for studying this problem was to investigate
the effect of quenched random impurities on the order of the
transition, as it is well known that such randomness can often force a
first-order transition into a continuous one \cite{b}. An interesting
example of this is the transition from the normal metal to the
Meissner phase which although first-order
for a pure sample \cite{hlm}, is driven to a continuous transition in
the presence of disorder \cite{bc}. Our analysis was based on the work
of BNT. In the course of our work, we became aware that the
representation of the FRG used by BNT might not always be sensitive to
the existence of fixed points and therefore their conclusions on the
nature of the transition should be re-examined.

The results from our analysis for the pure system are, however,
qualitatively the same as those of BNT: no stable fixed-point exists
in the FRG, which may be considered as evidence of a first-order
transition. The inclusion of quenched random impurities changes this
result dramatically. We find that the FRG supports a stable fixed
point if the coupling functions (to be defined shortly) are restricted
to the space of functions with non-zero support only at reciprocal
lattice vectors $\lbrace {\bf G} \rbrace $ which correspond to the
Abrikosov lattice.

The starting point of the analysis is the Landau-Ginzburg free energy
functional for type-II superconductors in the presence of quenched
random impurities
\begin{eqnarray}
\label{e1}
\nonumber F = \int \ d^{d}r \Biggl \lbrack & & {1\over 2m^*}|(\nabla +
ie^{*}{\bf A}) \psi _{i} |^{2} +(\tau + \delta \tau({\bf r}) )|\psi
_{i}|^{2} \\ & + & g_{0} |\psi _{i}|^{2} |\psi_{j}|^2 + {1\over
  2\mu_{0}}(\nabla \times {\bf A} - {\bf H})^{2} \Biggr \rbrack
\end{eqnarray}
where $\lbrace \psi _{i}\rbrace$ are a set of $n/2$ complex fields
(the implicitly repeated indices $i$ and $j$ are to be summed from $1$
to $n/2$), ${\bf H}$ is the external magnetic field, and $\delta \tau$
is a field representing the quenched random impurities whose quenched
average is zero, and whose correlator is given by $\langle \delta
\tau({\bf r}) \delta \tau({\bf r}') \rangle = \Delta _{0} \delta ^{d}
({\bf r}-{\bf r}') $.  As defined the theory has two coupling
constants, $g_{0}$ and $\Delta _{0}$. We have written the theory for
arbitrary spatial dimensionality $d$. In the physical case of $d=3$,
the external field ${\bf H}$ picks out a transverse plane $(x,y)$ in
which the Abrikosov lattice is formed. In general $d$, we define the
field ${\bf H}$ to be directed in a $d-2$ hyperplane
${\bf r}_{\perp}$, so that the Abrikosov lattice is still confined
to the two-dimensional $(x,y)$ plane transverse to this field.

We shall first concentrate on the pure system ($\Delta _{0}=0$). Since
we are interested only in the critical region, there are simplifications
which may be made to the form of the free energy.  We refer the reader
to Ref.\cite{bnt} for the details. The key point is that one may
restrict ones attention to the lowest Landau level, which corresponds
to writing the order parameter in the form
$\psi _{i}(x,y,{\bf r}_{\perp}) = \phi _{i}(z,{\bf r}_{\perp})
\exp (-\mu ^{2}z^{*}z/4)$
where $z=x+iy$ and $\mu ^{2}=e^{*}H $. A perturbative analysis of
this simplified theory reveals that one-loop contributions diverge due
to critical fluctuations in the $(d-2)$ hyperplane , thus leading to
an upper critical dimension defined by $d_{c} - 2 = 4$. To control
these divergences, one implements a renormalization group procedure
(perturbatively in the parameter $\epsilon = 6-d$) and, as found by
BNT, the theory is {\em not} closed under renormalization; i.e. the
theory defined with a simple coupling constant $g_{0}$ is not a
sufficiently general theory. To cure this problem, BNT generalized the
coupling constant to a coupling {\em function}, which under the
influence of translational invariance leads one to consider the new
bare theory:
\begin{eqnarray}
\label{e2}
\nonumber
F & = & \int \ d^{d-2}r_{\perp} \int dz \ dz^{*} \ e^{-\mu^{2}|z|^{2}/2}
(|\nabla _{\perp}\phi _{i} |^{2}+\tau |\phi _{i}|^{2}) \\
\nonumber
& + & \int d^{d-2}r_{\perp} \int dz_{1} dz_{1}^{*} \ e^{-\mu^{2}|z_{1}|^{2}/2}
\int dz_{2} dz_{2}^{*} \ e^{-\mu^{2}|z_{2}|^{2}/2} \\
& & \ \ \ \ \ \times
g(|z_{1}-z_{2}|^{2}) \
|\phi _{i}(z_{1},{\bf r}_{\perp})|^{2} \
|\phi _{j}(z_{2},{\bf r}_{\perp})|^{2} .
\end{eqnarray}
To close the FRG flow equations, BNT then made a representation of the
coupling function $g(u)$ of the form $g(u) = {1\over 2}\int
_{0}^{\infty} {dx\over x} \ \rho(x) e^{-(\mu^{2}/4x)u}$ leading to a
(complicated) FRG flow equation for the `weight' $\rho(x)$. Numerical
integration of this flow equation failed to lead to a fixed-point form
of $\rho(x)$ and this was interpreted by BNT as signalling a
first-order transition.

One point we wish to make in this Letter is that representing the
coupling function in terms of $\rho(x)$ is rather unfortunate. One
can show that stable FRG fixed points exist for the model described by
Eq.(\ref{e2}) for the cases $n \rightarrow \infty$ \cite{lr} and
$n=0$.  In each case, the weight function corresponding to the
fixed-point form of the coupling function is ill-defined. In other
words, a failure to find a fixed point form for the weight $\rho(x)$
is not necessarily
related to the existence or otherwise of a fixed point form for
$g(u)$. Details of this rather technical point will be given in a future
 publication \cite{mn}, together with the full derivation of the results
 given below.  For now, we shall simply indicate how
to proceed in an alternative fashion. Instead of using the integral
representation of $g(u)$ defined above, one may construct a FRG flow
equation for the Fourier transform (FT) ${\tilde g}({\bf k})$ of the
coupling function (we stress that ${\bf k}$ is a {\em two-dimensional}
wave-vector.) It then turns out to be more convenient to
concentrate on the function ${\tilde f}({\bf
  k})=e^{-k^{2}/2\mu^{2}}{\tilde g}({\bf k})$.
By studying the two-point correlation function one then finds that the
Hartree and exchange diagrams (whose sum gives the one-loop correction
to the correlation exponent) are given respectively by ${\tilde f}(0)$
and $f(0)$ (where $f$ is the inverse FT of ${\tilde f}$) indicating that
this choice of representation is a natural one.

As mentioned above, one can show the existence of stable FRG fixed
points for the special limits of $n \rightarrow \infty$ and $n=0$. We
shall briefly describe these two cases. In terms of the coupling
function ${\tilde f}$ one may derive the following FRG flow equation
\begin{equation}
\label{a3}
\partial _{l}{\tilde f}({\bf k}) = \epsilon {\tilde f} - (n/2){\tilde
  f}^{2} - 2{\tilde f}\circ{\tilde f} - 2{\tilde f} f^{*}
\end{equation}
which is in a differential form, with $e^l$ corresponding to the scale change
of the FRG. We have introduced the notation:
\begin{equation}
\label{a4}
{\tilde \alpha}\circ{\tilde \beta} \equiv \int {d^{2}p\over 2\pi} \
{\tilde \alpha} ({\bf p}){\tilde \beta}({\bf k}-{\bf p}) \cos
^{2}({\bf p}\times {\bf k}/2)
\end{equation}
and
\begin{equation}
\label{a4p}
\alpha^{*}({\bf k}) \equiv \int {d^{2}p\over 2\pi} \ {\tilde \alpha}
({\bf p}) \cos ({\bf p}\times {\bf k})
\end{equation}
with the definition ${\bf p}\times {\bf k} = p_{x}k_{y}-p_{y}k_{x}$ (we
have also scaled wave-vectors by $\mu$, and ${\tilde f}$ by $1/(2\mu)$.)

For $n \rightarrow \infty$ the flow equation is trivially solved. We have
$\partial _{l}{\tilde f} = \epsilon {\tilde f}-{\tilde f}^{2}$
(for the rescaled function ${\tilde f} \rightarrow 2{\tilde f}/n$),
with stable fixed-point solution ${\tilde f} = \epsilon$. This
solution indicates that the coupling `function' $g(u)$ has the nature
of a distribution \cite{lr} (leading to problems of definition for the
weight function of BNT.) One may consider $1/n$ corrections to this
result by analysing the full flow equation, Eq.(\ref{a3}). One sees
that a simple analytic expansion in powers of $1/n$ is not possible,
so the limit of $n \rightarrow \infty$ must be seen as somewhat
singular.  The other extreme of $n=0$ may also be considered in a
simple way.  Setting $n=0$ in Eq.(\ref{a3}), indicates that one may
obtain the fixed point ${\tilde f}({\bf k}) = (\epsilon \pi /2)
\delta^{2}({\bf k})$. A stability analysis shows that this
fixed point is stable against arbitrary perturbations.
This fixed point corresponds to $g(u)= {\rm
  const.}$ (which again leads to problems in the definition of the
weight function of BNT.) It is noteworthy that a naive expansion
around this result in powers of $n$ is not possible - the case of
$n=0$ is therefore also singular. An interesting possibility is to
expand about the $n=0$ limit by making a scaling Ansatz of the form
${\tilde f} = (1/n)s({\bf k}/n^{1/2})$, where the scaling function $s$
is to be determined from the FRG. Such a strategy has so far yielded
no useful results, and a full discussion will therefore be
postponed until a future publication \cite{mn}.

Unfortunately the flow equation for ${\tilde f}$ for the physical case
of $n=2$ is not analytically tractable and one must resort to
numerical integration. This task is highly non-trivial as the flow
equation has the form of a non-linear integro-differential equation.
We have failed to find any {\em stable} fixed points for this flow
equation for finite $n>0$, although we have been able to generate many
{\em unstable} fixed points by means of a Newton root-finding scheme
applied to the fixed-point form of the flow equation.  Although we
cannot say with certainty that stable fixed points do not exist, we
are inclined to believe there are none. This `runaway' of the coupling
function signals a breakdown of perturbation theory which may be
interpreted as evidence of a first-order transition. We shall present
an alternative scenario for $n>4$ later in this Letter.

We now turn to the first main point of this Letter - the effect of
disorder on the transition. One may proceed through all the steps that
led to the model described in Eq.(\ref{e2}), but now with the addition
of the disorder field. [One uses the usual tricks to average over the
disorder \cite{lub}, thus producing an effective quartic interaction
with strength (-$\Delta _{0}$).] In a completely analagous fashion to
the pure case, one quickly finds that the theory with simple coupling
constant $\Delta _{0}$ is not closed under renormalization, and one is
therefore obliged to generalize the theory by replacing $\Delta _{0}$
with a function $\Delta (|z_{1}-z_{2}|^{2})$. We choose to make the
natural FT representation as before. We therefore have flow equations
for the functions ${\tilde f}({\bf k})$ and ${\tilde D}({\bf k}) =
e^{-k^{2}/2\mu^{2}}{\tilde \Delta} ({\bf k})$ of the form
\begin{eqnarray}
\label{e3}
\nonumber \partial _{l}{\tilde f}({\bf k})& = & \epsilon {\tilde f} -
(n/2){\tilde f}^{2} - 2{\tilde f}\circ{\tilde f} - 2{\tilde f} f^{*} +
4{\tilde f}\circ{\tilde D} + 2{\tilde f} D^{*} \\ & & \\ \nonumber \partial
_{l}{\tilde D}({\bf k})& = & \epsilon {\tilde D} - n{\tilde D}{\tilde
  f} + 2{\tilde D}\circ{\tilde D} + 2{\tilde D} D^{*} - 2{\tilde D} f^{*}
\end{eqnarray}
with the same notation as used above.

We investigated these coupled flow equations numerically for the physically
interesting case of $n=2$ and found no stable fixed points, but
only a runaway of the coupling functions. Again, this
breakdown of perturbation theory could be interpreted as signalling
 a first-order transition.  In the remainder of this Letter we shall maintain
 an alternative
scenario regarding this breakdown of perturbation theory which  leads  to
very interesting consequences.

We shall  interpret this breakdown of perturbation theory as a flow to
some strong-coupling fixed-point. In terms of the model representation
in Eq.\ref{e2}, this fixed-point is not accessible via an
$\epsilon$-expansion.  However, we may envisage constructing a ``$\phi
^{4}$'' model which is perturbatively close to this fixed-point. One
would expect that any model which is a candidate for describing the
vortex liquid - vortex crystal phase transition would be able to
describe the two broken symmetries of the low-temperature Abrikosov
phase. One broken symmetry is ODLRO, or phase coherence in the
directions ${\bf r}_{\perp}$, and the other broken symmetry is
the breaking of full rotational and translational symmetry
of the liquid phase down to those of the triangular Abrikosov
lattice. A model which can capture these broken symmetries arises if one
considers
the same theory as before but with the added constraint of restricting
${\tilde f}$ and ${\tilde D}$ to have non-zero weight only at values of
$k$ coincident with the set of reciprocal lattice vectors (RLV)
$\lbrace {\bf G} \rbrace$ of the  Abrikosov lattice.  In practice we set
${\tilde f}({\bf k}) = \sum _{\bf G} A({\bf G})\delta ^{2}({\bf
  k}-{\bf G})$ and ${\tilde D}({\bf k}) = \sum _{\bf G} B({\bf
  G})\delta ^{2}({\bf k}-{\bf G})$.  With these forms one
sees that the new field theory is closed under renormalization;
coupling functions not of this
form are not produced perturbatively.
Furthermore, the flow equations given above are
still valid. Of course such a model cannot
describe the liquid phase. In that respect it is similar to some of the
theories  of two-dimensional melting which, being
expressed in terms of Burgers' vectors, can only in a staightforward
way describe the
crystalline phase (see e.g.\cite{nh}). While we have been unable to derive this
RLV model starting from Eq.(\ref{e2}), we feel it captures the essence
of the symmetries broken in the transition and hence because of
`universality' arguments provides a way of calculating critical
exponents etc. at the transition, should it be a continuous one.
Another noteworthy point is that the FRG flow equations in
Eqs.(\ref{e3}) determine ${\tilde f(|{\bf k|})}$, and it is hard to see
how the rotational symmetry can be broken down to that of a triangular
lattice unless such breaking is introduced `by hand', as indeed is
done within our model. The introduction of the RLV model is the second
major point of our Letter.

One finds that the flow equations are immensely simplified by our
particular choice of the RLV - the RLV of the triangular Abrikosov
lattice.  For this choice the trigonometric terms in the convolution-type
integrals become equal to unity, and the functions ${\tilde f}$ and
${\tilde D}$ become self-reciprocal under the FT.  Proceeding with
this choice of the RLV, it is possible to make analytic progress with
the simple Ansatz $A({\bf G}) = A$, $B({\bf G}) = B$, i.e.  choosing
the RLV coefficients to be independent of ${\bf G}$.

For this Ansatz one easily finds fixed-point values for $A$ and $B$
given by $A_{FP} = \epsilon/(2(n-1)\delta^{2}({\bf 0}))$ and $B_{FP} =
(4-n)\epsilon /(8(n-1)\delta^{2}({\bf 0}))$ which can be proven to be stable
under arbitrary RLV perturbations for $1<n<4$.
The symbol $\delta^{2} ({\bf 0})$
is proportional to a free sum over RLV, or equivalently to the system
size. The one-loop correction to the correlation length exponent is
given by $2-1/\nu =(n/2){\tilde f}(0) + f(0) + D(0)$, and we
therefore have at the above fixed point
\begin{equation}
\label{e5}
\nu ^{-1} = 2 - {3n\over 8(n-1)}\epsilon + O(\epsilon ^{2}).
\end{equation}
To address the question of stability
of this fixed point, we consider perturbations to the coupling functions.
In order to remain within the space of coupling functions
consistent with the RLV model, the perturbations must in
turn be restricted to the RLV.
Explicitly, we write
\begin{eqnarray}
\label{pert}
\nonumber
{\tilde f} & = & {\tilde f}_{FP} + \sum _{\bf G} a({\bf G})
\delta ^{2}({\bf k}-{\bf G}) \\
{\tilde D} & = & {\tilde D}_{FP} + \sum _{\bf G} b({\bf G})
\delta ^{2}({\bf k}-{\bf G}) .
\end{eqnarray}
Substituting these forms into the flow equations Eqs. (\ref{e3}) and then
performing a linear stability analysis yields the eigenvalue
spectrum for the (infinite dimensional) stability matrix. Stability is
ensured by requiring that this spectrum has no positive eigenvalues. It
turns out that the spectrum is characterised by only two different eigenvalues,
a non-degenerate eigenvalue
$\lambda _{1} = -\epsilon $ and a highly degenerate eigenvalue
$\lambda _{2} = \epsilon (n-4)/(4(n-1))$
(to leading order in $\epsilon $),
which clearly shows the stability of this fixed point in the range $1<n<4$.
Interestingly, the value of the correlation length exponent and
the value of the stability eigenvalues (which are related to
correction-to-scaling exponents) are the same as those obtained
for the simple $O(n)$ model in the presence of disorder \cite{lub}
(but in two dimensions higher; note that in this Letter $\epsilon = 6-d$.)

One may also use this RLV Ansatz in the absence of disorder, i.e.
Eq.(\ref{a3}). In this case one finds a fixed-point
solution $A_{FP} = 2\epsilon /((n+8)\delta^{2}({\bf 0}))$ which is found to be
stable to arbitrary RLV perturbations only for $n>4$. The value of
$\nu $ at this fixed-point for $n>4$ is identical with that
obtained from the pure $O(n)$ model \cite{on} (again in two higher
dimensions.) The stability analysis for the pure case reveals an eigenvalue
spectrum characterised by two different eigenvalues, with values
$\lambda _{1} = -\epsilon $ and $\lambda _{2} = \epsilon (4-n)/(4(n+8))$.
Clearly the fixed point is only stable for $n>4$, and one may therefore
not make the direct connection to the pure $O(n)$ model, since $n=4$ plays
no special role in that case. (Intriguingly, the value of the eigenvalues
for the pure superconductor are the same as those for the Heisenberg fixed
point in the disordered $O(n)$ model \cite{lub} .)

The results for $n = 2$ in the presence of disorder suggest that our
problem is in the same universality class as the disordered $O(n)$
model, but in two dimensions higher. Hence the lower critical dimension
of our problem would be expected to be $4$. That is, in the presence
of disorder, one would not expect there to be a phase transition in $d
= 3$ or $d = 2$. However, our perturbative treatment of the disorder is
appropriate only for weak disorder.  Strong disorder might drive the system
to a gauge glass phase.
However,the recent simulation results
of Bokil and Young \cite {bk} suggest that the lower
critical dimension of the gauge glass is greater than $3$. The old argument of
Larkin\cite{lar} suggests at the very least that disorder modifies the
nature of any transition below $4$ dimensions.

In the absence of disorder, we have been unable to find for $n = 2$
any stable fixed points even within the RLV model, and deduce that the
original conclusion of BNT that the transition becomes first order
below $6$ dimensions is likely to be correct. Our calculation gives no
information on the lower critical dimension of the pure case. One of
us has argued \cite{mm} that thermal excitation of  phase fluctuations
does not permit the simultaneous existence of ODLRO and the vortex lattice
for $d < 4$. However, while the RLV model certainly ensures that the
low-temperature phase (if it exists) is crystalline, it does not necessarily
require that phase to have ODLRO.

Hence, it is very tempting to compare our predictions -- that in the absence
of disorder a first order transition should exist and that in the presence
of (weak) disorder no transition is to be expected -- with the
experimental results
of Fendrich et al.\cite{fen}, who found that they could suppress
the apparent first order transition seen in untwinned single crystals
of YBCO by introducing point defects via electron irradiation.
Futhermore, the resistivity in the disordered system did not show the
behaviour expected according to the gauge glass phase transition
hypothesis.  To our eyes their resistivity curves after irradiation
look consistent with a gradual pinning of the vortices as the
temperature is lowered, i.e. with the absence of a phase transition.

It is a pleasure to thank Alan Bray, Joonhyun Yeo and Martin Zirnbauer for
interesting discussions. Our special thanks go to Leo Radzihovsky who
participated in the early stages of this investigation. TJN
acknowledges financial support from SERC and SFB 341.

\end{document}